# Mega Regional Heat Patterns in US Urban Corridors

**Babak J.Fard,[1,2] Udit Bhatia,[2,3] Auroop R. Ganguly[2]**

[1]Department of Environmental, Agricultural, and Occupational Health, College of Public Health, University of Nebraska Medical Center, Omaha, NE

[2]Sustainability & Data Sciences Laboratory, Civil & Environmental Engineering, Northeastern University, Boston, MA, USA.
[3]Civil and Environmental Engineering, Indian Institute of Technology, Gandhinagar, 48 Gujarat, India.

*Corresponding Author Email: a.ganguly@northeastern.edu

**Abstract**

Current literature suggests that urban heat-islands and their consequences are intensifying under climate change and urbanization. Here we explore the relatively unexplored hypothesis that emerging urban corridors (UCs) spawn megaregions of intense heat which are evident from observations. A delineation of the eleven United States UCs relative to their underlying climatological regions (non-UCs) suggest a surprisingly mixed trend. Medians and trends of winter temperatures over the last 60-years are generally higher in the UCs but no such general trends are observed in the summer. Heat wave metrics related to public health, energy demand and relative intensity do not exhibit significantly higher overall trends. Temperature and heat wave indices in the UCs exhibit high correlations with each other including across seasons. Spatiotemporal patterns in population, along with urbanization, agriculture and elevation, exhibit high (positive or negative) rank correlations with warming and heatwave intensification. The findings can inform climate adaptation in megalopolises.

## Introduction

The current consensus in the literature points to regional impacts of global warming and changes in land use as the dominant causes for intensification of temperature and heat waves at local to regional scales (*1–3*). In cities around the world, urban heat islands (UHIs) are considered among the major concerns for public health (*4*, *5*), energy demand (*6*, *7*) and decaying infrastructure (*8*), leading to adaptation and mitigation strategies for climate resilience (*9*, *10*) . Studies based on observations and model simulations have suggested intensification of UHIs over the last several decades resulting from the combined effects of anthropogenic climate change and intense urbanization including growing urban sprawl (*11–13*) . Previous studies have examined urban heat and their consequences across multiple scales from relatively close-knit communities and buildings within cities to megacities, finding evidence for intensifying UHIs (*14*, *15*).

Recent studies on urban resilience or sustainable urban systems have increasingly described the concept of interconnected city-of-cities (*16*). These are urban corridors (UCs) where human and engineered, and even certain natural systems have gotten so closely connected (and are getting even more intimately connected over time) that the entire corridor of multiple cities may need to be considered as a single coupled system (*17*, *18*). These UCs have also been called emerging megaregions (especially over the continental United States) or megalopolises (across the world) (*19*, *20*). However, to our knowledge, there has been no concerted effort in the literature or best practice to examine the impacts of these growing urbanized megaregions on temperature and heat waves relative to the non-urbanized regions in the climate region(s)(*21*) (CRs) in which they belong (hereinafter, non-UCs). No attempt has been made to understand the relation among temperature-based indices in the UCs or their relationships with potential explanatory variables.

Here we analyze daily temperature data, specifically average maximum and minimum temperatures, to examine the hypothesis that regional warming and heat waves have both been (statistically) significantly intensified in the megaregions represented by the urban corridors when compared with the background non-urban regions in identical climate zone(s). We analyze median temperature means, minima and maxima, as well as three heat wave metrics, specifically, cooling degree days, nighttime minima over three consecutive 24-hour periods, and 95th percentiles, along with their trends from 1956-2015, over winter (December-January-February: DJF) and summer (June-July-August: JJA) at each of the eleven emerging US megaregions or UCs relative to the non-urban regions in identical climate zones (non-UCs). Furthermore, we examine the hypotheses that impact-relevant metrics for warming and heatwave intensification correlate highly with each other, as well as with potential explanatory variables such as the degree of urbanization, spatiotemporal patterns of population in the UCs and non-UCs, along with agriculture and topography. Rank correlations among the variables are compared within the UCs as well as based on the differences between the UCs and the corresponding non-UCs.

## Materials and Methods

### Experimental Design

To investigate the potential effect of UCs on temperature we have used the geographical boundaries developed and suggested by Regional plan Association (RPA) (*17*). These boundaries show the expected development of US megaregions by 2050. RPA has used population density and growth, projected employment growth rate, and the integration between nucleolus high populated counties with their adjacent highly integrated counties to quantitatively distinguish these boundaries. A group of experts from RPA have then used the professional judgement and satellite imageries and finalized the map of 11 US megaregions by 2050. For the purpose of this study, we compare the trend and central tendencies of several temperature and heatwave variables for each UC and its corresponding non-UC for a 60-year study period. The 9 climate zones along with 11 UCs are used to group the observed data into 11 UC/non-UC pairs, each containing similar climatic characteristics.

Here, we study the chronic effect of UCs on temperature during Summer (JJA) and winter (DJF) using Tmin, Tavg, and Tmax values from GHCN-Monthly version 3(*22*) for 1956-2015. For calculating the acute effects on temperature, we extracted daily maximum and minimum temperature values from GHCN-D version 3(*23*), and used them to calculate the following three heatwave measures as in the following equations:

| | | |
|---|---|---|
| 1 | Mean annual three-night consecutive warmest nighttime minima event. (CNM) | (Critical to understand impact of heat waves on human health (*24*)) |
| 2 | Cooling degree days: $CDD = \max\left(0, \frac{t_{max}+t_{min}}{2} - 23.89\right)$ | (Driver of energy consumption(*25*)) |
| 3 | $95^{th}$ percentile of annual extremes of near-surface temperature (P95) (*26*). | |

Each station from the monthly or daily dataset is inside a UC or non-UC group. For each UC and study variable, we compare the significance level of regional trends with the corresponding non-UC. Also compare with its corresponding non-UC, the central tendency of annual averages of each variable over each UC through the 60-year study period. We then test the associations between the variables, within themselves and with the potential explanatory variables. Populations and their change in UCs, also their difference with non-UCs, the percent of urban areas and croplands, and the average elevation and the elevation range are considered as explanatory variables for further investigation. Population data are extracted from American Community Survey (ACS) 5-year estimate 2011-2015(*27*) (Table S3). Population changes are calculated for 1990-2015 from National Historical Geographic Information System (NHGIS) (Table S3). Percent urban and percent cropland are extracted from USDA land use database (*28*).

### Data Preparation

Use of discontinuous data in trend analysis can lead into unreliable conclusion (*29*). Regression-kriging models are shown to work best for predicting unavailable data with spatial distribution (*30*). Here we use Geographically Weighted Regression (GWR) (*31*) for each time step to fill the missing data. For monthly temperatures, stations with more than 10% missing values or more than one year of consecutive missing values are eliminated. Then Geographically Weighted Regression (GWR) is used to impute missing values in each time step. Elevation of stations are used as the independent variable for training the GWR model. Residual ordinary kriging is then used to add the predicted residuals for unavailable data points from the residuals of the regression model from observed values (*32*). After data preparation, monthly values in each station are averaged into JJA and DJF for each year. For daily temperature values, stations with less than 719 months of recorded data which ended before 2014 were eliminated. Furthermore, the stations with more than 20% of missing values during summer, or more than 30 consecutive days of missing values were deleted.
As the last step, we use Linearly Weighted Moving Average (LWMA) method to replace remaining missing values. For a consecutive n day of missing values, we calculate LWMA for 2n days before and LWMA for 2n days after the missing period and average them into the missing values. The data preparation is therefore completed with six seasonal temperature values for each monthly station and each year, also three heatwave values for each daily station and each year. ACS population data and NHGIS data are added from county level into each UC or non-UC group (Table S3). The land type percentages and the elevation variables are averaged over each climate zone and used for their corresponding UC/non-UC pairs.

### Statistical Analysis

For all the statistical tests in this study the p-value of 0.05 is considered as the threshold for the significance of the test. To distinguish the potential effect of each UC on study variables, we first test if there is a significant trend for each station. For this purpose, nonparametric Mann-Kendall is used (*33*).To eliminate the effect of serial dependence in the time series data, the regional Mann-Kendall test is used (*34*). To compare the regional trends of each UC and corresponding non-UC, first a field significance test is applied on each group of stations using False Discovery Rate (FDR) (*35*, *36*). This method adjusts p-values for the stations in each group (increases the values.). After the adjustment any group with at least one significant trend is considered as regionally significant. We have furthermore calculated the percentage of stations with significant adjusted trends for each scenario and used test of equal proportions (*37*) to compare these proportions in each pair of UC/non-UC.

For comparing the central tendencies, the Wilcoxon rank-sum test is applied to each UC and non-UC pair to test if the medians of the annual averages of each study variable in each group are significantly different. The Spearman rank correlation is used to assesses the potential relation among the study metrics, also with the potential explanatory variables. Population data is extracted from ACS five-year estimate of 2015 in county level to calculate the population in each of the UC and non-UC areas. To calculate the population changes data used from IPMUS National Historical Geographic Information System (*38*), for years 1990-2015.

## Results

### Megaregions and Climate Zones

Megaregions or urban corridors (UCs) are networks of metropolitan regions which form large geographic clusters of cities that are joined together by five layers of relationships, specifically, environmental systems (including topography), infrastructure systems, economic links, settlement and land use patterns, as well as shared culture and history, which together define common interests and hence forms bases for unified policy decisions (*18*). Eleven such UCs have been identified in the continental United States (*17*, *39*) (Fig.1.). While certain megaregions have been identified decades ago (e.g., NE was spotted by geographer Jean Gottman in 1961 (*40*)), they have all been growing and intensifying in their relationships, and a few are relatively recent. Further details on the eleven UCs, including current population, current GDP contributions and current and projected growth, and related information, are in the online supplement (SI: Table S1).

The National Centers for Environmental Information scientists have identified nine climatologically consistent climate regions (CRs) within the continuous or conterminous United States (CONUS) (*21*) , which have been found useful to put current climate patterns into a historical context. Figure 1(a) depicts these regions and their names in nine distinguishable colors encompassing CONUS map. Further details on CRs in the CONUS, including location, geography, climate, basic indicators of land-use and agriculture intensity, are in the online supplement (SI: Table S2).

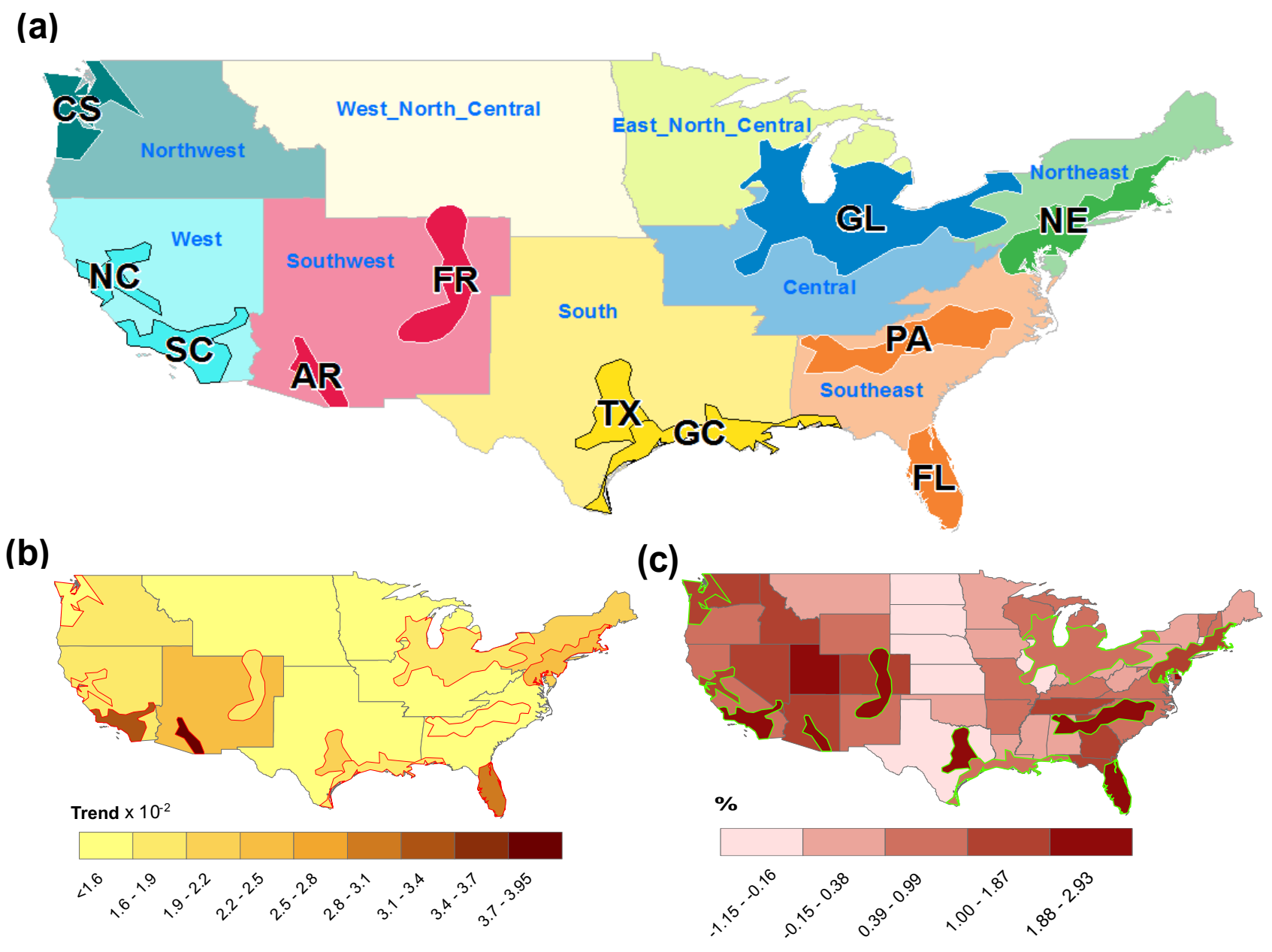

**Fig. 1. US Urban Corridors (UCs) and their temperature and population trends.** (a) 11 US UCs (*17*), and their corresponding non-Urban (non-UC) areas distinguished by the same colors but different shades, clipped out of the nine US Climate Regions (*21*). (b) trends of annual average of Tavg in UCs and non-UCs for 1956-2015. (c) Average of annual population growth in different states and UCs (1990-2015).

The megaregions or UCs and the hydroclimate regions or CRs within the conterminous United States or CONUS were delineated for different purposes and by different groups. To our knowledge, this may be one of the first studies that attempt to bring these delineations together to generate specific insights. Figure 1 visualizes the UCs and CRs within one map and shows a relevant climate trend (trends in average temperature over the last ~60-years) as well as a relevant population trend (growth of annual population in states and megaregions over the last ~25-years). Figure 1a shows that each UC lies within a specific CR, however there are three relative minor exceptions (GL, GC, FR) where the UCs jut out slightly into a different CR. Figure 1b depicts (via visual inspection: statistical analysis and interpretation follow in later figures) that in terms of 60-year trends in average temperatures, three UCs (AR, SC, FL) show distinguishably higher values relative to their background CRs, four UCs (NE, GL, TX, GC) show slightly higher values than their CRs, while two (NC, PA) are comparable but two UCs (CS, FR) exhibit lower trends than the CRs to which they belong. Figure 1c suggests from visual inspection that while almost all eleven UCs exhibited higher population growth relative to the background non-UC portion of the states, the relative difference is not so marked for three UCs (GL and perhaps CS and FL).

Visual inspection suggests that while relative behavior of megaregions versus non-megaregions appears (more or less) intuitive in terms of population growth, average temperature trends are much less so. Only three of eleven UCs clearly show relative intensification in the UCs, while two exhibit lower values.

## Temperature Differences and Trends

Figure 2 shows the 60-year median difference between the eleven United States UCs (or megaregions) and the surrounding regions (non-UCs or "non-megaregions") within the same climate regions(s) (CRs), as well as the level of regional significance of the 60-year trends in the UCs and non-UCs, for maximum (Tmax), average (Tavg) and minimum (Tmin) temperatures in the Summer (JJA) and Winter (DJF).

Based on the median UC versus non-UC temperatures shown in Figure 1(a), the eleven UCs appear to self-organize in three prime clusters and one megaregion (SC) with blended characteristic of its two neighboring clusters: UCs in the northern and western US coasts forming one category (CS, NC and to a some extent, SC), the inland UCs with one exception forming a second category (GL, FR, PA), the four UCs at or near the US eastern seaboard forming one category with another at the US-Mexico border (FL, AR, GC, TX, NE), while SC may belong to the first category or even form a separate standalone category. This UC, standing between the first and last group also shows a behavior in the middle of them. Figure 1(b) shows these categories in different colors, where rough West Coast, Inland and East Coast groups are suggested, even though the delineations are not perfect (e.g., AR and TX cluster with the East Coast while SC only partially clusters with the West Coast). The East Coast cluster (this group includes AR and TX for the purposes of this discussion) almost always exhibits statistically significantly higher median values in the UCs relative to the corresponding non-UCs for all cases (i.e., Tmax, Tavg and Tmin for JJA and DJF), while the Inland cluster almost always exhibits statistically significantly lower median values in the UCs, and in the West Coast cluster the UCs appear to show statistically significantly higher median temperatures in the Winter (DJF) but mostly statistically lower median temperature in the Summer (JJA) relative to their non-UCs.

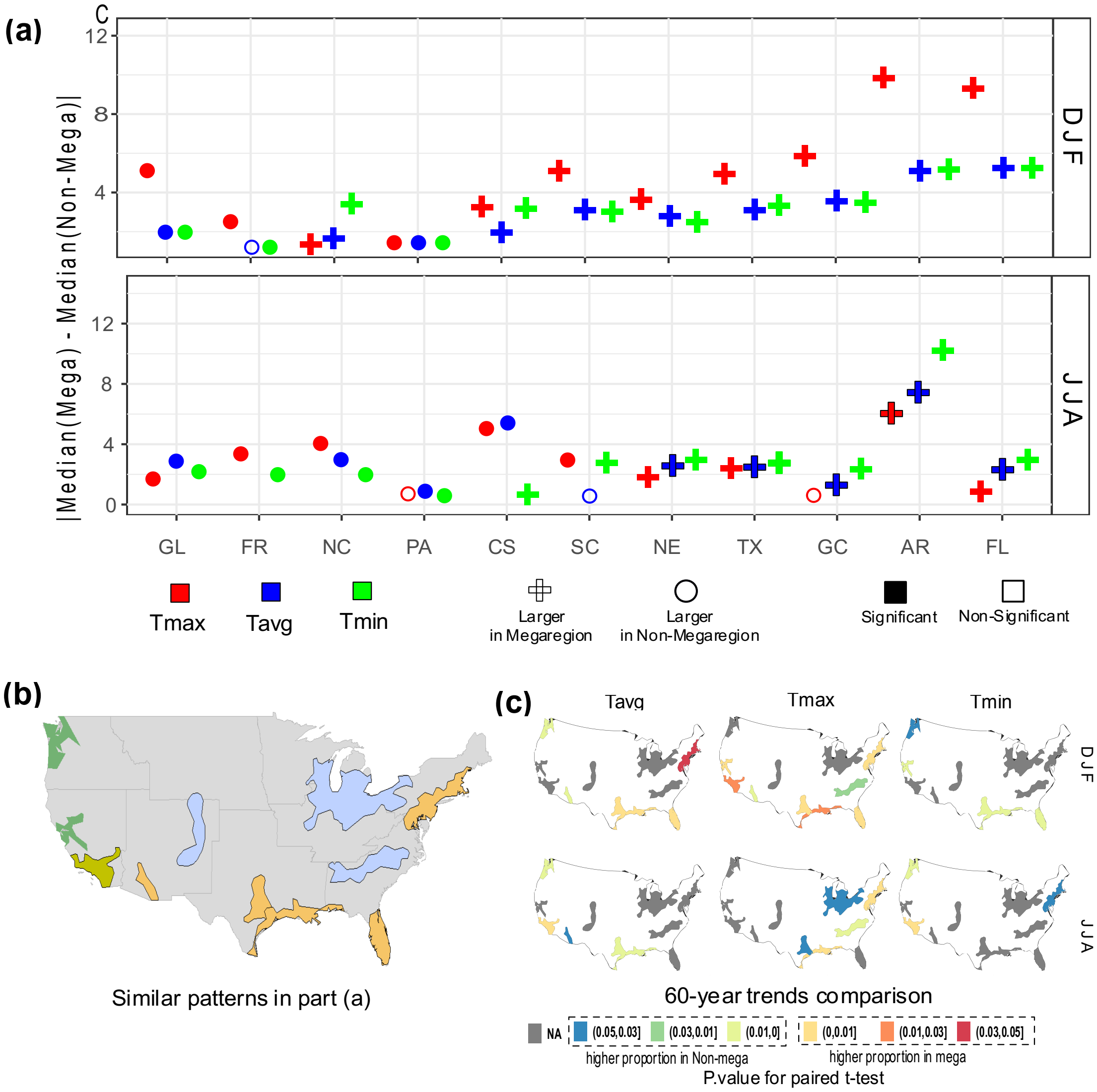

**Fig. 2. Comparisons of UCs vs. non-UCs for annual average values and trends for temperature variables.** (a) Differences in the medians of annual averages over UCs and non-UCs for the study period; for winter and summer. Detailed boxplots of annual average values are shown in Figures S1 to S3. (b) Suggested grouping of UCs based on the observed similar patterns in part (a). (c) results of the test of equal proportions between each UC and non-UC for the proportion of stations with significant trend, after FDR correction for trend p-values.

Figure 2(c) shows 60-year trends in temperatures (based on proportion of stations showing statistically significant trends after correction for false discovery rates): Cooler colors (light green, dark green and blue) indicate (progressively) higher trends in non-UCs relative to the UCs, while warmer colors (yellow, orange and red) indicate (progressively) higher trends in UCs relative to their non-UCs. Here we cannot observe any consistent clustering Pattern among UCs which are apparent in Figure 2(a). The Inland cluster displays no statistically significant pattern either way in most cases. Statistically significant higher trends in the UCs are observed in the average and maximum Winter temperatures in the Northeast, Florida, Texas and Gulf Coast (i.e., most of the UCs belonging to the East Coast cluster) but not in the minimum temperatures. The UCs in the West Coast cluster (NC and CS) either do not exhibit any significant trends or are mostly significantly lower than non-UCs, although the trend in SC is either not significant or higher than the corresponding non-UC. The seasonal contrast is marked: During the Summer, the UCs mostly (but not consistently) exhibit either no significant trends or statistically significantly higher trends, while in the Winter, the UCs mostly (but again, not consistently) exhibit either no significant trends or statistically significantly lower trends, than the corresponding non-UCs. The Northeast and Texas show an interesting pattern: statistically significant and relatively large higher trends in the UCs compared to the corresponding non-UCs in the Winter for Tavg and Tmax, along with often statistically significantly lower trends in the UCs for Tmin and Tavg in the Summer. The Great Lakes show no statistically significant difference in the UC versus non-UC regions, except in the Summer Tmax, when the UC trend in statistically significantly lower than the corresponding non-UC.

**Multi-Metric Analysis of Heat Waves**

Heat waves are compared in terms of absolute magnitude and trends between UCs and corresponding non-UCs in terms of three metrics (Figure 3). The first metric is based on Cooling Degree Days (CDD) and relates to energy demand, the second metric is based on the 95th percentile (P95) and relates to a measure of extreme intensification, while the third metric is based on consecutive (three-nights) nighttime minima (CNM) which is a measure of human (dis)comfort levels based on impacts on human physiology.

Figure 3a shows the differences between absolute heat waves in the UCs versus their non-UCs. The boxplots are sorted from left to right based on the differences in the median values for each UC and non-UC pair. Looking into each diagram three groups are distinguishable from left to right. Left group, those with significantly higher medians in non-UCs; middle group, those with non-significant differences, and, the right group; that contain pairs that UCs have significantly higher medians. CDD and CNM (top and bottom rows in Figure 3(a)) have almost equal size groups of first and last, with CDD has 5 pairs in the first, 1 in the second and 5 in the last group. CNM contains five pairs in the first group, zero in the second, and six in the last group. While the two major groups contain the same groupings that we found in Figure 2. (GL, FR, NC, PA, CS) are in the left group that perfectly match the groupings of west and inland UC from Figure 1. (FL, AR, GC, TX, NE) are the right-side group in this figure that again perfectly match the east coastal grouping in Figure 1. SC slides from group 1 to group 2 and also CDD is the only UC that has non-significant difference. This again perfectly matches with the SC grouping in figure 1. P95 contains 6 pairs in the left group, 3 in the middle group, and 2 in the right group. It shows some deviation from the other 2 diagrams, although the groupings are not dramatically changed.

Figure 3b shows a lack of consistent statistically significant trends in heat wave intensities of UCs versus non-UCs over the last 60-years. Statistically significantly higher 60-year trends are observed in the UCs (versus the corresponding non-UCs) in terms of CDD for GC (Gulf Coast) and Florida (FL) and in terms of CNM for NE (Northeast), GC and CS (Cascadia), while statistically significantly lower regional level of 60-year trends in the UCs are observed in terms of P95 in NC (North California) and PA (Piedmont Atlantic).

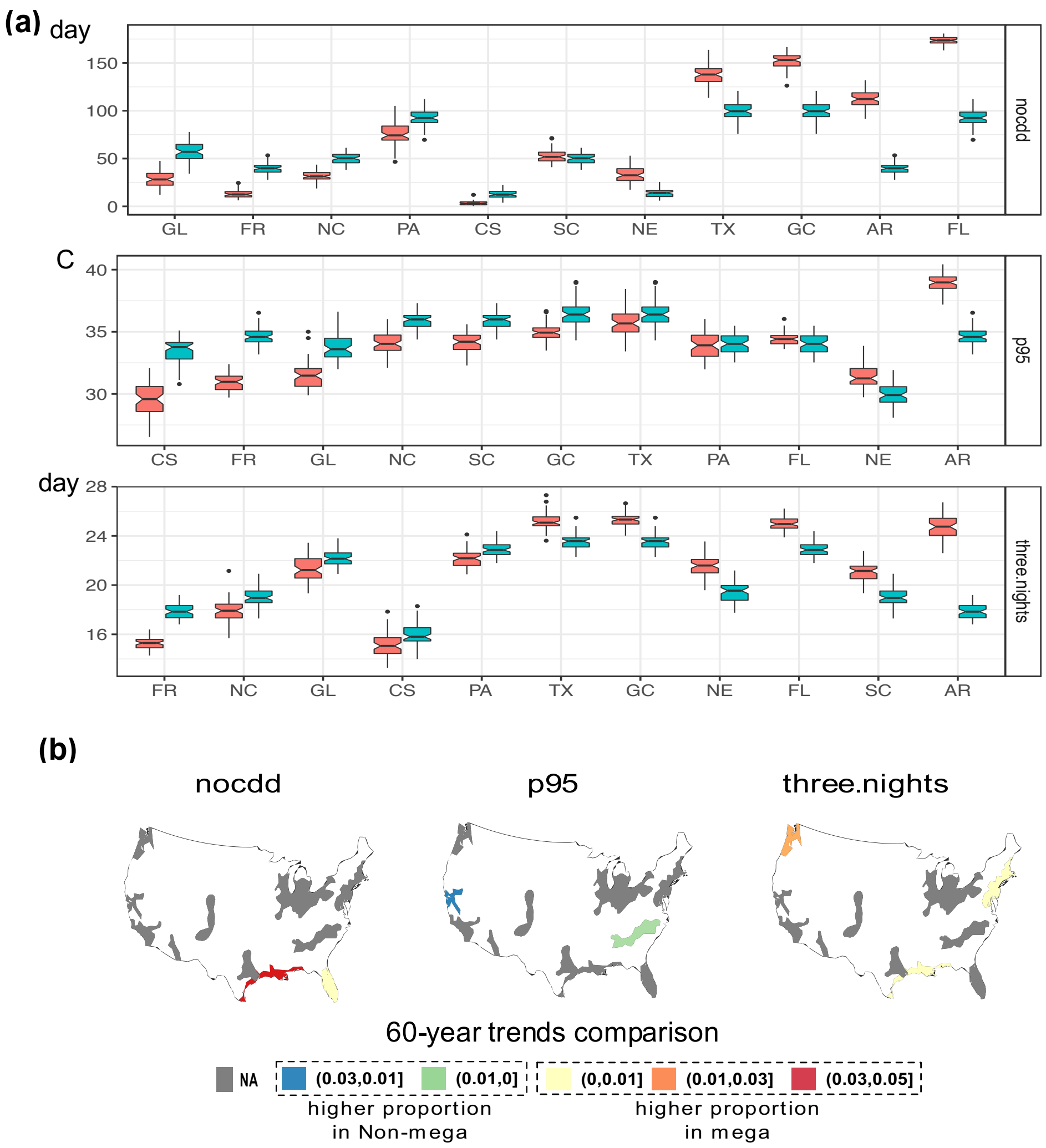

**Fig. 3. Annual average values and trend comparisons for Heatwave Indicators. (a)** Boxplots of annual average heatwave indicators over each UC and non-UC for the study period. The notches indicate 0.95 Confidence Interval around each median value. Greens represent UC and reds show non-UCs. **(b)** results of test of equal proportions of stations with significant trends in each UC vs. non-UC, after corrections for regional effect using False Discovery Rate (FDR).

## Spearman's Rank Correlations

The nine temperature and heat wave indices studied here are the minimum, average, and maximum temperature in winter (DJF) and summer (JJA), as well as three heat wave metrics: one related to public health (three consecutive nighttime minima temperature above a threshold), energy demand (number of cooling degree days), and a dynamic threshold (the 95% percentile). Temperature and heat wave indices are computed for the UCs and for the difference between the UCs and corresponding non-UCs. Eight variables with presumed explanatory power, specifically percent urbanization (within UCs), percent cropland (within non-UCs), average elevation and elevation span of the climate region, as well as four variables related to population: absolute value of population in the UCs, difference in population in UCs relative to the corresponding non-UCs, percent change in population in the UCs, and percent change in the difference of population between UCs relative to the corresponding non-UCs.

Figure 4 shows the rank correlations among temperature and heat wave indices with each other as well as their rank correlations with the eight potential explanatory variables. Figure 4a (left) shows the temperature and heat wave indices in the UCs while Figure 4b (right) shows the same but for the differences in temperature and heat waves indices between the UCs and their corresponding non-UCs.

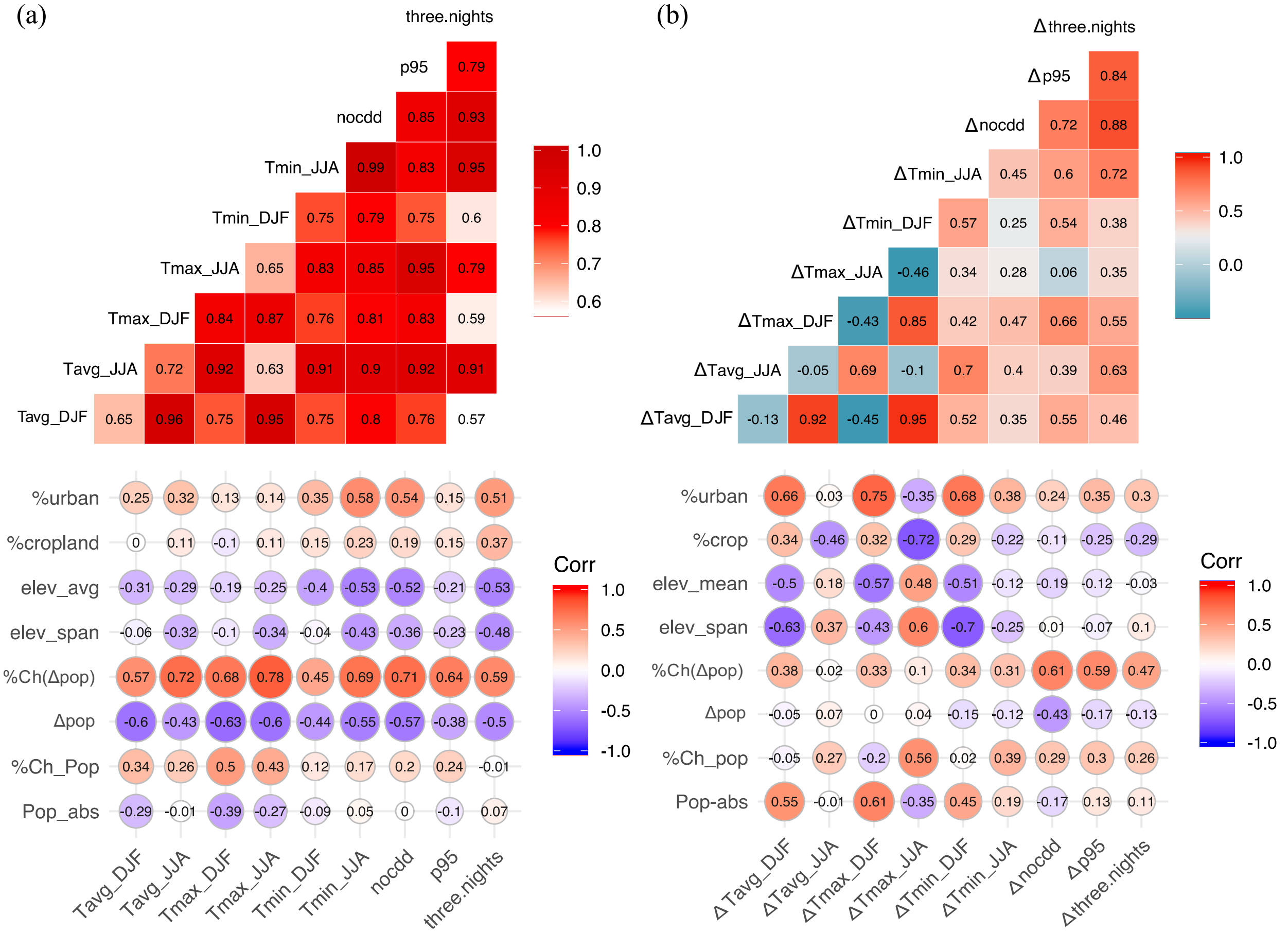

**Fig. 4. Rank Correlations between study parameters and descriptor variables over 11 UCs.** Top part of each diagram shows the results of Spearman rank correlation between study parameters (temperature and heatwave values). The bottom parts show rank correlations between study parameters and potential descriptor variables. % urban and %cropland are percent of urban and cropland area calculated for corresponding Climate Region of each UC. elev_avg and elev_span are the average of elevation and the elevation span for the corresponding climate regions. %Ch(Δpop) is the average annual percentage change for the UC-(non-UC) populations. Δpop is the UC-non-UC population for ACS 5-year estimation of 2015. %Ch_Pop is the average annual percentage change in UC population and Pop_abs represent the population in UC based on the ACS 5-year estimation of 2015. **(a)** The temperature variables for UCs are used in correlation. **(b)** The differences of temperature variables, (UC-(non_UC)), are used in correlation.

Figure 4(a) shows that all nine indices of temperatures and heat waves correlate well with each other for all seasons over the UCs. The nine temperature and heat wave indices in the UCs exhibit a relatively high rank correlation with each other. Of the eight variables with potential explanatory power, the rank correlations are the highest (either in a positive or a negative direction) with the spatiotemporal gradients of the population differential (i.e., difference of the population in UCs versus the corresponding non-UCs. The population differential by itself reflects a spatial gradient (between UCs and corresponding non-UCs), while the percent changes (of the population differential) reflect temporal changes potentially related to further urbanization of the UCs. The highest negative rank correlations (of temperatures and heat waves within the UCs) are with the population differential between the UCs and the corresponding non-UCs. The positive rank correlation may show the impact of the surrounding non-UCs on the temperatures and heat waves of the UCs. Thus, if the population differential is less, then the non-UCs are expected to resemble the UCs more, and such relatively more urbanized non-UCs are likely to have a relatively higher warming influence (since the dissipation of heat from the UCs to the surrounding non-UCs may reduce) on the UCs they surround. However, the highest positive correlations are with percent changes in the UCs versus non-UCs population differential. The positive changes in population differential are likely a result of further urbanization of the UCs (or vice versa) relative to the non-UCs, which in turn is expected to increase the urban heat effect. Other than the UC to non-UC population differential, the other variables with percent urbanization exhibits a positive rank correlation (which likely reflects the intensification of heat from the urban environment) while average elevation shows a negative rank correlation (which most likely is just indicative of the cooling effect of higher elevation).

Figure 4(b) may look deceptively identical to Fig. 4(a) in terms of the variables for which rank correlations are shown, but there is one important difference. The nine variables pertaining to temperatures and heat waves are not for the UCs themselves, rather these are for the differences between the UCs and the corresponding non-UCs. The differences in rank correlations in the two sets of figures, viz, Fig. 4(a) and 4(b), are rather dramatic. Comparing the numbers at the top showing the temperature or heat waves, the strength of the rank correlations reduces dramatically, with some variables exhibiting negative rank correlations in summer versus winter. Comparing the numbers at the bottom, the spatiotemporal patterns in population differential (UCs minus non-UC) do not correlate (or anticorrelate) as highly. However, other variables rise in prominence, such as the positive rank correlations with both urbanization and the percent of non-UC land marked as cropland, and the negative rank correlations with elevation. While the positive rank correlations with degree of urbanization is intuitive, the one with percent cropland may be owing to added heat from crops (e.g. the implication of “corn sweat”.

Further comparing Fig. 4(a). and Fig. 4(b), we may realize that on part (a) the changes are one dimensional. Meaning that moving on each of the horizonal or vertical lines of part (a) the changes in rank correlations are not dramatic. It means that the relationship of a variable to other variables is almost constant. But in part (b) the color blending is not linear but creates almost four quadrants in bottom section of Figure 4(b). among them upper left quadrant clearly shows the highest density of both highest positive and negative values. That quadrant represents the relationships between land characteristics and parameters of chronic changes in temperature. Based on that we can conclude that among the interaction of four sets of variables; the two-explanatory set of population parameters and land characteristics; the two sets of gradual temperatures and heatwaves; land characteristics have high effects on the gradient of chronic changes in temperature between UCs and non-UCs. Also, on the top part of Fig. 4(b) it is clearly seen that all long-term temps have positive correlations for the same season and negative for opposite seasons.

## Discussion

The world's population is getting concentrated in a few large cities. In developed economies like the US, the extent and pace of urbanization has been rapid. Furthermore, large cities have grown regionally interconnected in terms of movement of peoples and commodities, financial institutions and social communities, as well as businesses, resources and lifeline infrastructures such as transportation, energy, communication and water (*19*). Thus, large megalopolises or urban corridors, comprising a group of interconnected city-of-cities, have emerged. Meanwhile, even as growing contributions to greenhouse gas emissions and land use changes have increased urban contributions to climate change, the world's cities with disproportionately large populations within relatively small areas are often facing the brunt of the consequences. Urban heat islands have been intensifying resulting in human lives lost and misery, along with impacts on public health, infrastructures resilience and energy consumption. Based on these considerations, it would appear logical to examine the hypothesis that large megaregions of intense heat have emerged or are already beginning to emerge in the urban corridors of the continental United States (CONUS). However, what is surprising from our study, to a first order, is that lack of any consistent signal in this regard. When examined further, especially at the level of individual urban corridors in the CONUS, the spatiotemporal patterns are revealed to be complex. When temperature and heat wave indices are correlated with themselves and with variables of interest (including potentially confounding variables), new insights emerge many of which appear explainable upon hindsight while some lead to new research questions and hypotheses.

Throughout this study, we have looked into the potential regional effects of UCs on chronic and acute changes of temperatures. For this purpose, we have compared each UC with its neighboring non-UC within the same Climate Region. A clear geographical distinction in the difference of average values is clear for both acute and chronic measures. It can be hypothesized and be further investigated as the effect of large-scale climatic patterns that can still be the major deriving reason for these patterns. The absence of a similar grouping pattern in trend differences can strengthen this hypothesis, considering that climatic patterns have not changed for the timeline of study. Although, looking into individual UCs and their differences in trends based on the season or variable can lead into interesting questions and insights that are worth more detailed studies.

In the quest for potential descriptors, we realize that changing the study variables from UC temperatures into their differences with non-UCs can drastically change the orders and magnitudes of rank correlations. If one is comparing various UCs for their difference in temperature measures, without considering their non-UCs, the population difference with their non-UCs and the rate of this change is of paramount importance. But when these same regions are being compared based on their temperature difference with their non-UCs, the land use and topology of their climate region plays higher importance.

The quality, quantity and resolution of observed data, especially deeper into the past, are major concerns which limit observation-based insights and design of experiments. Urban corridors are often still emerging or even beginning to emerge, thus the dynamic nature of the geographical extent of these corridors and the concentration of people and built environment may limit the insights. Urban corridors are not homogeneous or consistently urban, while areas designated as non-urban have a mix of semi-urban, rural, agricultural, forest and barren land, among others. In addition, the sample size of urban corridors is still rather small to always draw statistically meaningful conclusion. Thus, insights and findings need to be carefully caveated. Simulated data, whether from weather reconstruction such as reanalysis, or archived weather and climate model archives, may offer additional and more consistent insights, but suffer from limitations inherent in such models. The influence of potentially important and confounding variables needs to be examined further through hypothesis-driven studies, with both field and simulated based testbeds.

**Acknowledgments**

This work was supported by four National Science Foundation Projects, including NSF BIG DATA under Grant 1447587, NSF Expedition in Computing under Grant 1029711, NSF CyberSEES under Grant 1442728, and NSF CRISP type II under Grant 1735505. The bulk of the work was completed while the first author was a PhD student at Northeastern University.

## Supplementary Materials

| Megaregion | Population 2010 | % of US Population | Population 2025 | Population 2050 | Projected Growth (2010-2050) | GDP (2005) billion | % of US GDP |
|---|---|---|---|---|---|---|---|
| Arizona Sun Corridor | 5,653,766 | 2% | 7,764,211 | 12,319,771 | 117.90% | $191.04 | 2% |
| Cascadia | 8,367,519 | 3% | 8,748,143 | 11,864,378 | 41.80% | $337.40 | 3% |
| Florida | 17,272,595 | 6% | 21,449,652 | 31,122,998 | 80.20% | $608.08 | 5% |
| Front Range | 5,467,633 | 2% | 6,924,457 | 10,222,370 | 87.00% | $229.20 | 2% |
| Great Lakes | 55,525,296 | 18% | 60,678,100 | 71,263,185 | 28.30% | $2,073.87 | 17% |
| Gulf Coast | 13,414,934 | 4% | 16,334,987 | 23,666,122 | 76.40% | $524 | 4% |
| Northeast | 52,332,123 | 17% | 58.4 million | 70.8 million | 35.20% | $ 2920 (2010) | 20% |
| Northern California | 14,037,605 | 5% | 16,350,872 | 21,159,995 | 50.70% | $623 | 5% |
| Piedmont Atlantic | 17,611,162 | 6% | 21,687,449 | 31,342,393 | 78.00% | $486 | 4% |
| Southern California | 24,361,642 | 8% | 29,010,560 | 39,381,675 | 61.70% | $1,036 | 7% |
| Texas Triangle | 19,728,244 | 6% | 24,809,567 | 38,132,600 | 93.30% | $ 817 | 7% |

Table S1. Eleven UCs, including current population, current GDP contributions and current and projected growth (*17*)).

| Regions and States | Cropland | Grassland pasture and range | Forest-use land | Special-use areas | Urban areas | Miscellaneous other land | Total land area (1,000 acres) |
|---|---|---|---|---|---|---|---|
| Northeast | 11.00% | 4.42% | 59.10% | 10.34% | 12.00% | 3.14% | 111,123 |
| East North Central | 42.53% | 7.17% | 34.84% | 7.58% | 3.38% | 4.50% | 157,556 |
| Central | 38.59% | 12.69% | 35.59% | 5.69% | 5.99% | 1.44% | 195,652 |
| Northwest | 11.48% | 31.60% | 37.59% | 11.66% | 1.66% | 6.01% | 156,856 |
| South | 24.57% | 42.68% | 21.75% | 4.17% | 2.84% | 3.99% | 354,399 |
| Southeast | 10.82% | 8.38% | 61.10% | 7.87% | 8.93% | 2.90% | 179,169 |
| Southwest | 5.66% | 60.33% | 19.03% | 11.96% | 1.35% | 1.68% | 269,251 |
| West | 5.97% | 46.48% | 13.18% | 21.28% | 3.42% | 9.66% | 169,958 |
| West North Central | 29.25% | 52.40% | 9.83% | 6.25% | 0.32% | 1.95% | 297,136 |

Table S2. Major uses of land, by Climate region (*21*), aggregated from included states, 2012 (*28*)

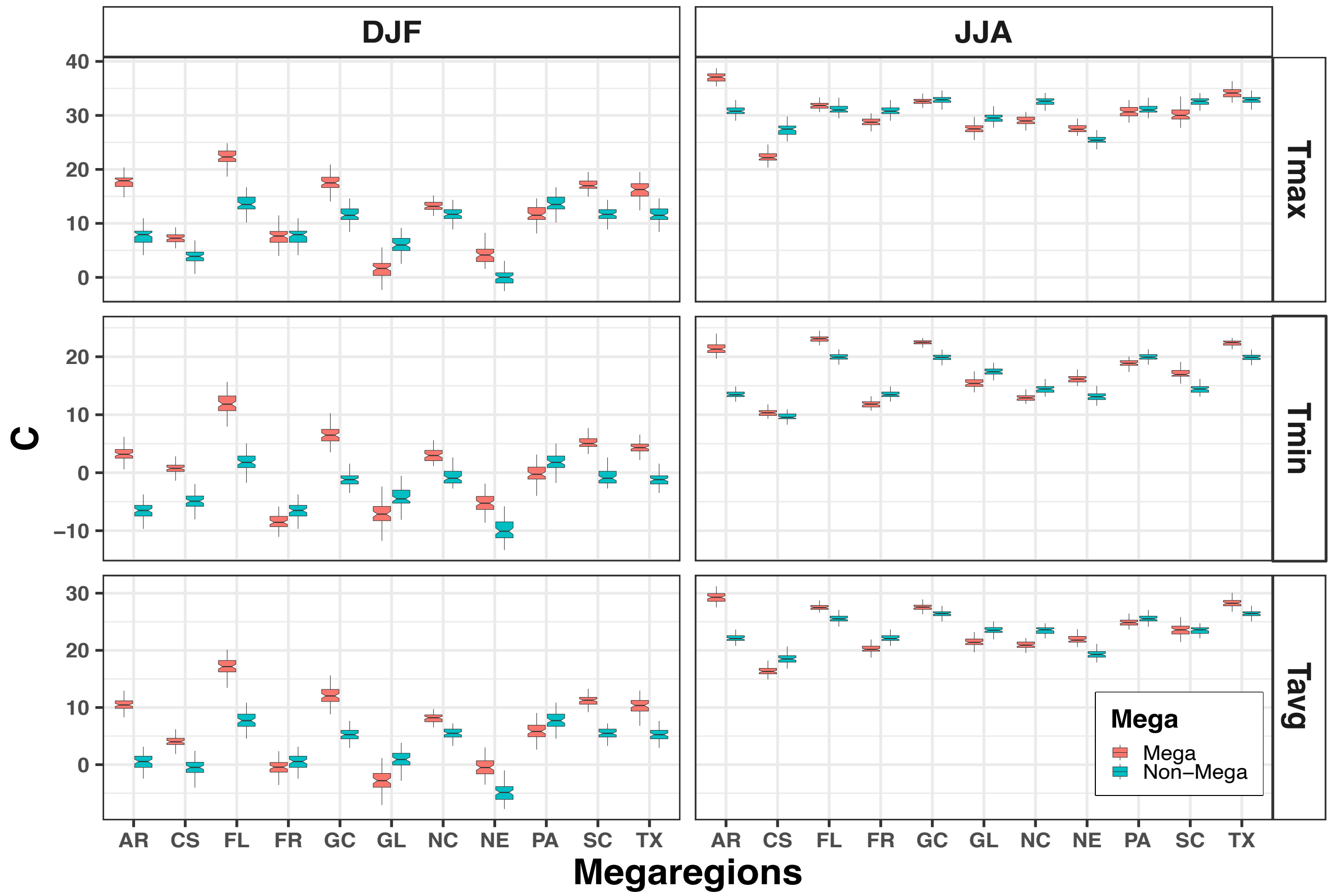

Fig S1. Tmax for DJF/JJA

| | | Population | | | |
|---|---|---|---|---|---|
| Megaregion | Symbol | Absolute | % Change (average annual) | Absolute (Mega-Nonmega) | % Change Average Annual (Mega-nonMega) |
| Arizona Sun Corridor | AR | 5,889,740 | 2.85 | 501,093 | 1.38 |
| Cascadia | CS | 8,685,125 | 1.66 | 4,829,006 | 0.35 |
| Florida | FL | 18,075,029 | 2.51 | -479,522 | 1.548 |
| Front Range | FR | 5,725,374 | 2.55 | 336,727 | 1.08 |
| Great Lakes | GL | 55,968,405 | 0.764 | 39,993,903 | 0.18 |
| Gulf Coast | GC | 7,280,017 | 0.95 | -9,680,820 | 0.8819 |
| Northeast | NE | 51,511,922 | 1.07 | 42,136,094 | 0.812 |
| Northern California | NC | 14,535,519 | 1.64 | 12,977,489 | 0.671 |
| Piedmont Atlantic | PA | 18,247,386 | 2.1 | -307,165 | 1.138 |
| Southern California | SC | 25,126,551 | 1.96 | 23,568,521 | 0.991 |
| Texas Triangle | TX | 20,940,466 | 1.92 | 3,979,629 | 1.8519 |

Table S3. Population parameters calculated for spearman rank correlation analysis. Absolute population aggregated from Census county level data (*27*). The historical values to calculate average changes are extracted from IPMUS National Historical Geographic Information System (*38*), for years 1990-2015

| | |
|---|---|
| ACS | American Community Survey |
| AR | Arizona Sun Corridor (Megaregion) |
| CDD | no. of Cooling Degree Days |
| CNM | Consecutive I(three-night) Nighttime Minima |
| CR | Climate Region |
| CS | Cascadia (Megaregion) |
| DJF | December-January-February |
| FDR | False Discovery Rate |
| FL | Florida (Megaregion) |
| FR | Front Range (Megaregion) |
| GC | Gulf Coast (Megaregion) |
| GDP | Gross Domestic Product |
| GHCN | Global Historical Climate Network |
| GL | Great Lakes (Megaregion) |
| GWR | Geographically Weighted Regression |
| JJA | June-July-August |
| LWMA | Linearly Weighted Moving Average |
| NC | Northern California (Megaregion) |
| NE | Northeast (Megaregion) |
| NHGIS | National Historical Geographic Information System |
| P95 | 95th percentile of annual extremes of near-surface temperature |
| PA | Piedmont Atlantic (Megaregion) |
| RPA | Regional plan Association |
| SC | Southern California (Megaregion) |
| TX | Texas Triangle (Megaregion) |
| UC | Urban Corridor |
| UHI | Urban Heat Island |

Table S4. Abbreviations